\documentclass[superscriptaddress,aps,prd,nofootinbib,twocolumn,showpacs]{revtex4-1}
\usepackage[english]{babel}
\usepackage[utf8]{inputenc}
\usepackage{float}
\usepackage{graphicx}
\usepackage[caption=false]{subfig}
\usepackage{color}

\usepackage{amsfonts,ulem}
\usepackage{amsmath}
\usepackage{amssymb}
\usepackage{graphicx}
\usepackage{mathrsfs}
\usepackage{multirow}

\newcommand{\ev}[1]{\ensuremath{\left\langle #1\right\rangle}}

\begin{document}

\title{Empirical constraints on the high-density equation of state \\from multi-messenger observables}

\author{Márcio Ferreira}
\email{marcio.ferreira@uc.pt}
\affiliation{CFisUC, Department of Physics,
University of Coimbra, P-3004 - 516  Coimbra, Portugal}

\author{M. Fortin}
\affiliation{N. Copernicus Astronomical Center, Polish Academy of Science,
Bartycka,18, 00-716 Warszawa, Poland}

\author{Tuhin Malik}
\affiliation{BITS-Pilani, Department of Physics, K.K. Birla Goa Campus, GOA - 403726, India}

\author{B. K. Agrawal}
\affiliation{Saha Institute of Nuclear physics, Kolkata 700064, India}
\affiliation{Homi Bhabha National Institute, Anushakti Nagar, Mumbai -
400094, India}

\author{Constan\c ca Provid\^encia}
\affiliation{CFisUC, Department of Physics,
University of Coimbra, P-3004 - 516  Coimbra, Portugal}
\date{\today} 

\begin{abstract} 
We search for possible correlations between neutron star observables
and thermodynamic quantities that characterize high density
nuclear matter. 
We generate a set of model-independent equations
of state  describing stellar matter from a Taylor expansion around
saturation density. Each equation of state which is a functional of the nuclear
matter parameters is thermodynamically consistent,
causal and compatible with astrophysical observations.
We find that the neutron
star  tidal deformability and radius are strongly correlated with the
pressure, the  energy density and the sound velocity at different
densities. Similar correlations are also exhibited by 
a large set of mean-field models based on non-relativistic and relativistic
nuclear energy density functionals.
These model independent correlations can be employed to constrain
the equation of state at different densities above saturation 
from measurements of NS properties with multi-messenger  observations. 
In particular, precise constraints on the radius of PSR
  J0030+0451 thanks to NICER observations would allow to better infer the properties of matter around two times the nuclear saturation density.

\end{abstract}

\maketitle

\section{Introduction}

The properties of the equation of state (EoS) of nuclear matter at
supra-saturation densities, that is above the nuclear saturation density 
 $n_0\sim 0.16$ fm$^{-3}$, still remain an open question in nuclear
physics. Neutron stars (NSs) are unique astrophysical objects
through which the properties of super-dense neutron-rich nuclear
matter  at zero temperature can be studied. Constraining the EoS
requires combining astrophysics and nuclear physics. Astrophysical 
observations are important probes for the dense nuclear matter
properties. Several NSs with a mass about 
two-solar masses detected during the last
decade set quite stringent constraints on EoS of
nuclear matter. The pulsar PSR J1614$-$2230 is,
among the most massive observed pulsars, the one with the smallest
uncertainty on the mass $M=1.906\pm0.016\, M_\odot$ 
\cite{Arzoumanian2017,Fonseca2017,Demorest2010} (masses are reported 
with $1\sigma$ error-bars or equivalently 68.3\% credibility 
intervals throughout this work). Other two pulsars with a mass above
two solar masses are PSR J0348$+$0432 with $M=2.01 \pm 0.04 M_\odot$
\cite{Antoniadis2013} and the recently detected MSP J0740$+$6620 
with a mass $2.14 {\scriptsize\begin{array}{c}+0.10\\-0.09\end{array}}M_\odot$ 
\cite{Cromartie2019}.\\

Detecting gravitational waves (GWs) emitted during the coalescence of binary 
NS systems is also one of the most promising way to probe high density 
behavior of the EoS for dense stellar matter.
The analysis of the compact binary inspiral event GW170817 has placed
upper bounds on the NS combined dimensionless tidal deformability
\cite{TheLIGOScientific:2017qsa}. Using a low-spin prior (consistent
with the observed NS population), the combined dimensionless
tidal deformability of the two NSs that merged during the event 
was determined to be  $\tilde{\Lambda}\le 800$ with 90\% confidence. 
A follow up reanalysis \cite{Abbott18} assuming the same EoS for the two NSs and for a spin 
range consistent with the one observed in Galactic binary NSs obtained
$\tilde{\Lambda}\le 900$ and  the tidal deformability of a $1.4$ solar mass NS
was estimated to be $70<\Lambda_{1.4M_{\odot}}<580$ at the 90\% level.
The detection of GWs from the GW170817  event was followed by the  electromagnetic
counterpart,  the gamma-ray burst (GRB) GRB170817A \cite{grb}, and the
electromagnetic transient AT2017gfo \cite{kilo}, that set extra
constraints on the lower limit of the tidal deformability
\cite{Radice2017,Radice2018,Bauswein2019,Coughlin2018,Wang2018}. This last
constraint  seems to rule out very soft
EoS: the lower limit of  the tidal deformability of  a 1.37$M_\odot$
star set by the above studies limits the  tidal deformability
 to  $\Lambda_{1.37M_\odot} > 210$ \cite{Bauswein2019}, 300 \cite{Radice2018},
  279 \cite{Coughlin2018}, and 309 \cite{Wang2018}.\\

In Ref. \cite{Lattimer01} Lattimer and Prakash have empirically observed  that for densities between 1.5$n_0$
and $2-3n_0$, the radius of the star scales with $p^{1/4}$ with $p$ the
pressure at these densities. This means that knowing the radius of a
NS with sufficient precision  will 
constrain the EoS of stellar matter  in this specific range of densities.  It
also raises the question whether other correlations between the
thermodynamic properties of nuclear matter in $\beta$-equilibrium and
NS properties such as the radius or tidal deformability could
exist. In this case further constraints on the NS EoS could be obtained
thanks to new measurements of the NS tidal deformability with
 future LIGO/Virgo detection of gravitational waves emitted from binary 
 NS mergers.
The precise determination of the radius of NSs, in addition to their mass,
 expected from the currently-operating NICER mission \cite{NICER},
and future X-ray observatories like the Athena X-ray telescope \citep{Athena} 
and eXTP \cite{eXTP} would also allow to constrain the EoS in various ranges
 of density. \\

NS properties such as the radius and mass  
can be obtained by solving 
the Tolman-Oppenheimer-Volkoff  (TOV) equations \cite{TOV1,TOV2} for a  static and
spherical star in
hydrostatic equilibrium, which requires the EoS as a input.  As a consequence a 
one-to-one correspondence  is established  between
the  NS mass and radius and the EoS of
$\beta$-equilibrated stellar matter. 
The possibility of inverting this mapping, allowing the determination of the
EoS from the measurement of the mass and radius of a large number of stars was discussed by
Lindblom \cite{Lindblom1993}.  
Later, it was proposed that a smaller  number of
astrophysical observations is required if  realistic EoS are
parametrized using  piecewise polytropes with transition densities from one polytrope to another chosen at well selected
densities \cite{Read2008}. A different approach was discussed  in
\cite{Ozel2009}, where it was shown that the  determination of the pressure at
three fiducial densities could be obtained from the measurement of
three different NSs.
\\

Correlations between nuclear matter parameters
and NS properties have been explored using several nuclear models \cite{Vidana2009,Xu2009,Ducoin2010,Ducoin2011,Newton2013,Alam2016,Malik2018,Carson:2018xri,Zhang2018}.
These studies, however, show a considerable model dependence 
since different models with similar values of the nuclear matter parameters
may result in different EoSs. In the present work, we study
the correlation of various astrophysical observables directly 
with the thermodynamical variables of the EoS to avoid the model 
dependence \cite{Tsang2019}.
The main objective of the present work is to look for
further correlations that could allow to establish constraints on the
EoS of nuclear matter from the observation of NS. We use  a large set of so-called meta-models that satisfy a
given number of well defined nuclear matter  and NS
properties \cite{Margueron2018a,Ferreira2019}.\\

The paper is organized as follows. In Sec. II, we introduce the EoS parametrization and  generating process for the meta-models. The correlation analysis on the generated set of EoS is developed in Sec. III. Finally, the conclusions are drawn in Sec IV.

\section{EoS parametrization}
We start from the generic functional form for the energy per particle of homogeneous nuclear matter 
\begin{equation}
{\cal E}(n,\delta)=e_{0}(n)+e_{sym}(n)\delta^2
\end{equation}
where $n=n_n+n_p$ is the baryonic density and $\delta=(n_n-n_p)/n$ is the asymmetry
with $n_n$ and $n_p$ being the neutron and proton densities, respectively.
This approach has been applied recently in several
works, \cite{Margueron2018a,Margueron2018b,Margueron2019}.
We consider a Taylor expansion of this energy functional around
  the saturation density $n_{sat}$ until fourth order as in \cite{Margueron2018a,Margueron2018b}:
\begin{align}
e_{0}(n)&=E_{sat}+\frac{1}{2}K_{sat}x^2+\frac{1}{6}Q_{sat}x^3+\frac{1}{24}Z_{sat}x^4\\
e_{sym}(n)&=E_{sym}+L_{sym}x+\frac{1}{2}K_{sym}x^2+\frac{1}{6}Q_{sym}x^3\\
&+\frac{1}{24}Z_{sym}x^4 \nonumber
\end{align}
where $x$ is defined as $x=(n-n_{sat})/(3n_{sat})$. The empirical parameters can be identified as the coefficients of the expansion.  The isoscalar empirical parameters are defined as proportional to successive density derivatives of $e_{0}(n)$,
\begin{equation}
 P_{IS}^{(k)}=(3n_{sat})^k\left.\frac{\partial^k e_{0}(n)}{\partial n^k}\right|_{\{\delta=0,n=n_{sat}\}},
\end{equation}
whereas the isovector parameters measure density derivatives of $e_{sym}(n)$,
\begin{equation}
 P_{IV}^{(k)}=(3n_{sat})^k\left.\frac{\partial^k e_{sym}(n)}{\partial n^k}\right|_{\{\delta=0,n=n_{sat}\}}.
\end{equation}
The corresponding empirical parameters are then 
\begin{equation}
 \{E_{sat},K_{sat},  Q_{sat},Z_{sat}\}  \rightarrow
  \{P_{IS}^{(0)},P_{IS}^{(2)},P_{IS}^{(3)},P_{IS}^{(4)}\}
\end{equation}
and
\begin{align*}
 &\{E_{sym},L_{sym}, K_{sym},  Q_{sym},Z_{sym}\}  \\
 &\rightarrow  \{P_{IV}^{(0)},P_{IV}^{(1)},P_{IV}^{(2)},P_{IV}^{(3)},P_{IV}^{(4)}\}.
\end{align*}

The coefficients of low orders are already quite well
  constrained experimentally \cite{Youngblood1999,Margueron2012,Li2013,Lattimer2013,Stone2014,OertelRMP16}, however  $Q_{sat},\, Z_{sat}$ and $K_{sym}, \, Q_{sym},
  \, Z_{sym}$ are only poorly known
 \cite{Farine1997,De2015,Mondal2016,Margueron2018b,Malik2018,Zhang2018,Li2019}. The saturation energy $E_{sat}$ and  saturation density $n_{sat}$ being rather well constrained, we fix their values throughout this work: $E_{sat}=-15.8$ MeV (the current estimated value is  $-15.8\pm0.3$ MeV \cite{Margueron2018a}), and $n_{sat}=0.155$ fm$^{-3}$.\\

With this approach, each meta-model is represented by a point in the 8-dimensional space of parameters.
Instead of analyzing the models on a fixed grid, we will employ random sampling of models through a multivariate Gaussian with zero covariance:
\begin{align*}
\text{EoS}_i & = \{E_{sym},L_{sym},K_{sat},K_{sym},Q_{sat},Q_{sym},Z_{sat},Z_{sym}\}_i \\
&\sim N(\boldsymbol{\mu},\boldsymbol{\Sigma})
\end{align*}
where the mean value vector and covariance matrix are, respectively, 
$$\boldsymbol{\mu}^T=(\overline{E}_{sym},\overline{L}_{sym},\overline{K}_{sat},\overline{K}_{sym},\overline{Q}_{sat},\overline{Q}_{sym},\overline{Z}_{sat},\overline{Z}_{sym})$$ 
and 
$$\boldsymbol{\Sigma}=diag(\sigma_{E_{sym}},...,\sigma_{Z_{sym}}).$$

In the present approach, as discussed in \cite{Margueron2018a}, no a-priori
correlations exist between the different parameters of the
EoS. However, as we will see, imposing experimental and
observational constraints will give rise to  correlations. The physical correlations among the empirical parameters arise from a set of physical constraints  \cite{Margueron2018a,Margueron2018b}. The parameters of the Gaussian distributions for each parameter are in Table~\ref{tab:initial_dist}.\\

\begin{table}[!htb]
  \begin{center}
  \centering
\setlength{\tabcolsep}{10pt}
\renewcommand{\arraystretch}{1.4}
\begin{tabular}{ccccc}
  \hline \hline 
\multirow{2}{*}{$P_i$} & \multicolumn{2}{c}{Initial dist.} & \multicolumn{2}{c}{Final dist.} \\ \cline{2-5} 
                  & $\overline{P}_{i}$ & $\sqrt{\sigma_{{P}_{i}}}$   & $\overline{P}_{i}$ & $\sqrt{\sigma_{{P}_{i}}}$ \\ \hline 
$K_{sat}$         & 230             & 20              & 233.35          & 18.24         \\
$Q_{sat}$         & 300             & 400             & 56.04           & 122.31        \\
$Z_{sat}$         & -500            & 1000            & -178.46         & 141.26        \\
$E_{sym}$         & 32              & 2               & 33.33           & 1.89          \\
$L_{sym}$         & 60              & 15              & 51.45           & 11.83         \\
$K_{sym}$         & -100            & 100             & -44.24          & 63.24         \\
$Q_{sym}$         & 0               & 400             & 237.52          & 299.42        \\
$Z_{sym}$         & -500            & 1000            & 372.98          & 698.72  \\ \hline \hline 
    \end{tabular}
    \caption{The mean $\overline{P}_{i}$ and standard deviation
$\sqrt{\sigma_{{P}_{i}}}$ of the multivariate Gaussian, where
$\sigma_{{P}_{i}}$ is the variance of the parameter $P_{i}$. 
Our EoSs are sampled using the initial distribution for 
$P_i$ assuming that there are no correlations
among the parameters. The final distribution for $P_i$ are obtained after imposing the 
filters as listed in the text. All the quantities are in units of MeV.
The values of $E_{sat}$ and $n_{sat}$ are fixed to $-15.8$ MeV and $0.155$ fm$^{-3}$,
respectively.}
\label{tab:initial_dist}
  \end{center}
\end{table}

We impose the following conditions to get a valid EoS: i)
be monotonically increasing (thermodynamic stability); ii) the
speed of sound must not exceed the speed of light (causality);
iii)  supports a maximum mass at least as high as $1.97M_{\odot}$
\cite{Arzoumanian2017,Fonseca2017,Demorest2010,Antoniadis2013}
(observational constraint); iv) predicts a tidal deformability
of $70<\Lambda_{1.4M_{\odot}}<580$ \cite{Abbott18} (observational
constraint); and v) the symmetry energy $e_{sym}(n)$ is positive. All the
EoS are in $\beta$-equilibrium.  We use the SLy4 EoS for the low density
region \cite{Douchin2001}.  A valid EoS must cross the SLy4 EoS in the
$P(\mu)$ plane below $n<0.10$ fm$^{-3}$ consistently with the range of core-crust transition densities for a large set of nuclear models \cite{Ducoin2011}.  The SLy4 EoS is matched with the
generated EoSs by requiring $P_{\text{SLy4}}(\mu)=P_{\text{EoS}}(\mu)$ with $\mu$ the chemical potential.

\section{Results}
In the present section we first discuss the properties of the set of EoS
we have  built after imposing the constraints listed above. Using these EoSs,
we then study correlations between NS observables and
the EoS properties at given densities.

\subsection{Empirical parameters values and NS properties}

After applying all the filters indicated above to $10^7$ sampled EoS,
we obtain 2121 valid EoS. This number is quite
small and it is mainly due to the constraint of causality
and the requirement that the generated EoS and the crust SLy4 EoS intersect in the $P-\mu$ plane. Using a simple interpolation between the crust and the core at some specific density is much less restrictive and the number
of valid EoS would be much larger. However, we consider it is important
to carry the information contained on the Taylor expansion not only to
supra-saturation densities but also to sub-saturation densities. \\

In Table \ref{tab:initial_dist}, the mean values and standard deviations
of the EoS parameters for the final distribution, after the constraints on the EoS were imposed,
are compared with the respective initial input.  It is interesting to
notice that well constrained parameters like $K_{sat}$ and $E_{sym}$,
and even $L_{sym}$, do not change much from the initial distribution,
while the parameters connected to the high orders, such as $Q_{sym}$
and $Z_{sym}$, converge to quite different mean values.

With this set of EoS,  we obtain the relation between the radius $R$ and the mass $M$  of the NS, solving the TOV
equations \cite{TOV1,TOV2}, and calculate the dimensionless tidal deformability $\Lambda$
\begin{equation}
\label{eq2}
\Lambda = \frac{2}{3}k_2\left(\frac{R}{M}\right)^5,
\end{equation}
where $k_2$ its  {quadrupole} tidal Love number, following Ref.~\cite{Hinderer2008}. 

\begin{figure}[!ht]
	\centering
	\includegraphics[width=1.0\columnwidth]{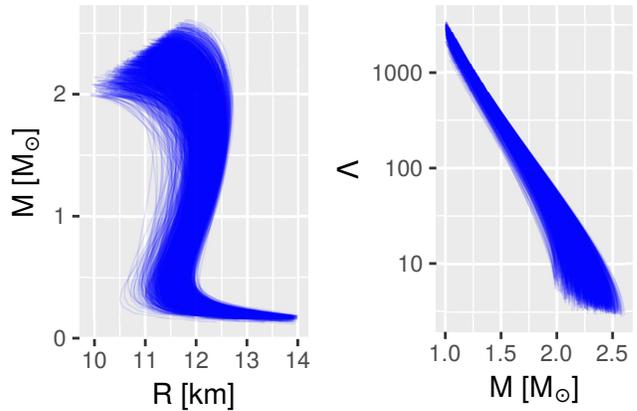}
	\caption{Mass vs. radius (left) and the mass vs. tidal deformability (right) diagrams for set of EoS built in the present study.}
\label{mrl} 
\end{figure}

In Fig.~\ref{mrl}, we plot the $M-R$ and the $\Lambda-M$
relations for the set of EoS.  In what follows, these results will be
used to study the correlations between NS observables and thermodynamic
quantities. As an example, we present in  Table~\ref{stat1} the mean value
and standard deviation for the tidal deformability  $\Lambda_{M_i}$ and
radius  $R_{M_i}$ of stars with masses $M_i=1.0,\, 1.2, \, 1.4, \, 1.6,
\, 1.8 ~M_\odot$. The results obtained for $M_{1.4}$ are well inside the
limits imposed by GW170817 \cite{Abbott18} for the tidal deformability
$70<\Lambda_{1.4 M_\odot}<580$ (but notice that this is also true 
without imposing maximum star mass of 1.97$M_\odot$) and $R=11.9\pm 1.4$
km.  On the other hand, if the constraints set by the electromagnetic
counterpart are also considered then our EoSs satisfies the lower limit
determined in \cite{Bauswein2019}, $\Lambda_{1.37 M_\odot}>210$, but  not
the limit calculated in  \cite{Radice2018,Coughlin2018,Wang2018}: 300, 279 and 309 respectively.
 However, in average and within a 95\%
confidence interval these lower constraints are all satisfied.  The set
of EoSs also satisfies the condition obtained for $R_{1.6 M_\odot}$
from an existing universal relation between the critical merger  remnant
mass to a prompt collapse  and the compactness of the maximum mass star,
i.e., $R_{1.6 M_\odot}\gtrsim 10.7$ km \cite{Bauswein2017,Koppel2019}.\\\\
The obtained results are also in agreement with \cite{Annala18},
where the maximum value $R_{1.4M_{\odot}}=13.6$ km and the minimum value
$\Lambda_{1.4M_{\odot}}=120$ were reported, using a generic family of EoS
that interpolate between chiral effective field theory results at low densities
and perturbative QCD at high densities. 
Furthermore, our results are compatible with \cite{Most18} (an extra condition on the allowed maximum NS mass was imposed, $M_{\text{max}}<2.16M_{\odot}$, though),
in which a mean value of $R_{1.4M_{\odot}}=12.39$ km and a $2\sigma$
confidence of $12.00<R_{1.4M_{\odot}}/\text{km}<13.45$ were determined using 
a piecewise polytrope parametrization of the EoS, which took into account nuclear matter calculations of the outer crust, near saturation densities, and perturbative QCD.

\begin{table}[!htb]
  \begin{center}
  \centering
\setlength{\tabcolsep}{8pt}
\renewcommand{\arraystretch}{1.1}
\begin{tabular}{ccccc}
\hline \hline 
               & mean    & std    & min     & max     \\ \hline 
$\Lambda_{1.0M_{\odot}}$ & 2967.88 & 283.78 & 1677.51 & 3597.70 \\
$\Lambda_{1.2M_{\odot}}$ & 1129.15 & 118.90 & 620.60  & 1377.93 \\
$\Lambda_{1.4M_{\odot}}$ & 467.53  & 57.89  & 243.53  & 579.93  \\
$\Lambda_{1.6M_{\odot}}$ & 201.96  & 31.36  & 93.31   & 267.29  \\
$\Lambda_{1.8M_{\odot}}$ & 87.54   & 18.48  & 29.41   & 126.26  \\ \\ 
$R_{1.0M_{\odot}}$       & 11.96   & 0.19   & 10.99   & 12.34   \\
$R_{1.2M_{\odot}}$       & 12.09   & 0.19   & 11.07   & 12.48   \\
$R_{1.4M_{\odot}}$       & 12.18   & 0.21   & 11.13   & 12.59   \\
$R_{1.6M_{\odot}}$       & 12.20   & 0.25   & 11.09   & 12.68   \\
$R_{1.8M_{\odot}}$       & 12.14   & 0.31   & 10.85   & 12.73 \\ \hline \hline 
\end{tabular}
\caption{Sample statistics for $\Lambda_{M_i}$ and $R_{M_i}$ (km): mean, standard deviation, maximum, and minimum values.}
\label{stat1}
  \end{center}
\end{table}

Interestingly, the minimum value obtained for $\Lambda_{1.4M_{\odot}}$ is 243.53 and, furthermore, only a very small percentage of the EoS failed to reproduce $\Lambda_{1.4M_{\odot}}<580$ (the maximum value  reached for $\Lambda_{1.4M_{\odot}}$ was $651.85$). In other words, the present set of EoS describes NSs 
with a narrow region of $\Lambda_{1.4M_{\odot}}$ with a mean value of 467.53, and  
fulfill $70<\Lambda_{1.4M_{\odot}}<580$ \cite{Abbott18}. 

\subsection{Correlation between thermodynamic quantities and NS observables}

\begin{figure}[!htb]
	\centering
	\includegraphics[width=1.0\columnwidth]{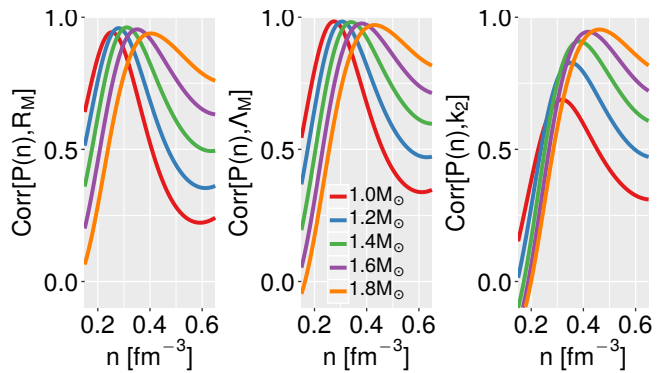}	
\caption{The density dependence of correlation coefficient of the pressure, $P(n)$ with radius $R$ (left), tidal deformability $\Lambda$ (middle), and Love number $k_2$ (right) for different NS masses as indicated obtained for Meta models.}
\label{fig:pressure}
\end{figure}

\begin{figure}[!htb]
	\centering
	\includegraphics[width=1.0\columnwidth]{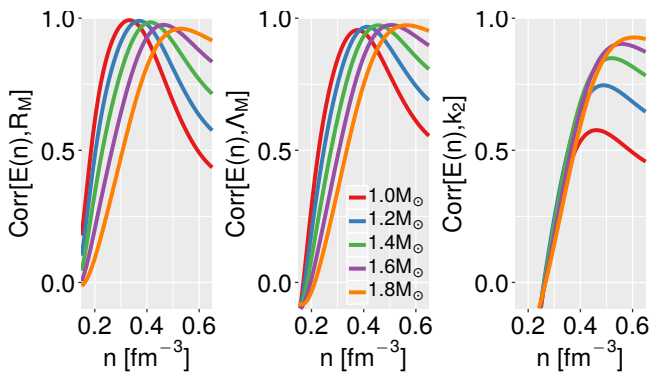}	
\caption{The density dependence of correlation coefficient of the  energy density, $E(n)$, with radius $R$ (left), tidal deformability $\Lambda$ (middle), and Love number $k_2$ (right) for different NS masses as indicated obtained for Meta models.}
\label{fig:energy}
\end{figure}

\begin{figure}[!htb]
	\centering
	\includegraphics[width=1.0\columnwidth]{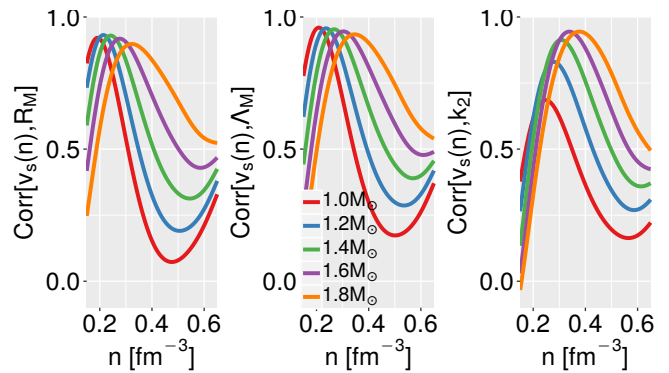}	
\caption{The density dependence of correlation coefficient of the  sound velocity, $v_s(n)$, with radius $R$ (left), tidal deformability $\Lambda$ (middle), and Love number $k_2$ (right) for different NS masses as indicated obtained for Meta models.}
\label{fig:speed}
\end{figure}

In the present section, we study the possible existing correlations between the thermodynamic
properties of dense stellar matter in $\beta$-equilibrium and NS
observables. In the following analysis we use the Pearson correlation coefficient  
$$ \text{Corr}\,[X,Y]=\frac{ \ev{(X-\mu _{X})(Y-\mu _{Y})}}{\sigma _{X}\sigma _{Y}},$$
where $\ev{...}$ is the expectation value and $\sigma_{X}$ and $\sigma_{Y}$ are the standard deviations of variables $X$ and $Y$, respectively. In particular, we consider
the correlation between the pressure in Fig.~\ref{fig:pressure}, the energy
density in  Fig.~\ref{fig:energy}, and the  speed of sound in  Fig.~\ref{fig:speed}, at each baryonic density with the radius, tidal deformability, and Love number of NSs
with $M=1.0,\, 1.2,\, 1.4,\, 1.6$ and $1.8~M_\odot$.

We first discuss the correlation of the NS properties with the
pressure as shown in Fig.~\ref{fig:pressure}. Interestingly one can identify strong correlations between the pressure
and the various NS properties that we considered, at certain densities. 
The left panel, for example, shows how the correlation between $P(n)$ and $R_M$ 
remains quite high in a relatively small range of densities for all the five masses considered.  
The maximum of the correlation shifts to larger densities, from $n\approx 0.25$ 
to 0.35~fm$^{-3}$, as the mass of the star increases. This is precisely the empirical
correlation identified by Lattimer and Prakash in
\cite{Lattimer01}. A similar and an even stronger correlation, with a coefficient very close to 1, is
observed between the pressure and the tidal deformability in the same range of densities (center panel). The Love number $k_2$ shows a quite strong
correlation at $n\approx 0.45$~fm$^{-3}$ but only for the larger masses (right panel).\\

The correlations between NS {observables} and the energy density are shown in Fig.~\ref{fig:energy}. Again, strong correlations are identified with all the NS {observables} but
at larger densities, $n\approx 0.32$ to 0.5 fm$^{-3}$ with
the smaller (larger) masses closer to the smaller (larger) value. 
Similar to the case of pressure, the correlation between the energy density and $\Lambda$ remains strong for all NS masses. The maximum correlation is obtained at higher densities when massive NS are considered.

Finally, for the speed of sound shown in Fig.~\ref{fig:speed} strong correlations are again present but occurring at smaller densities than
the ones obtained for the pressure, i.e. at densities $\approx 0.2 - 0.3$ fm$^{-3}$.

A strong correlation, i.e.,  $\text{Corr }\approx 1$, means that the sample variance of the NS observable is almost entirely linearly explained by the variance of the thermodynamical quantity. Therefore, one can use the NS observable as a way to constrain the thermodynamical quantities at different baryon densities. \\

\begin{figure}[!htb]
	\centering
	\includegraphics[width=1.0\columnwidth]{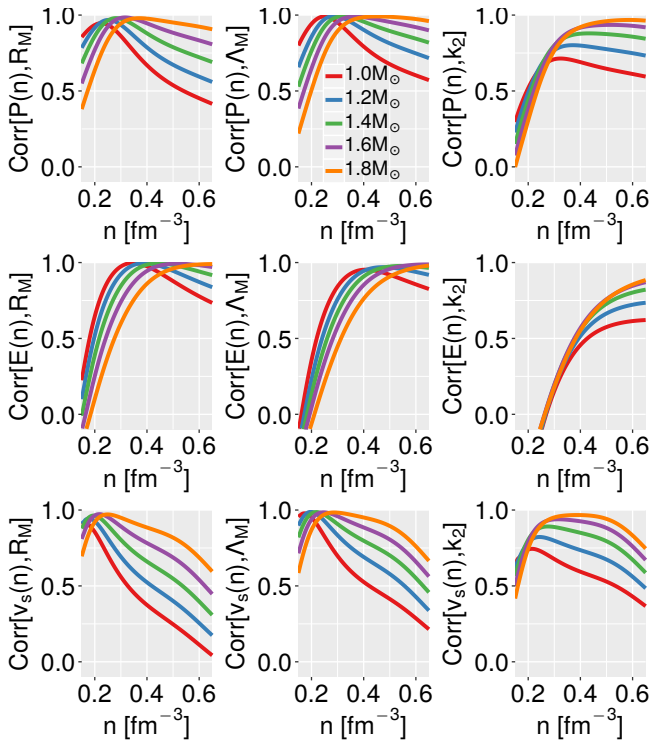}	
\caption{The density dependence of correlation coefficient of the  pressure $P(n)$ (top row), energy density $E(n)$ (middle row),
and sound velocity $v_s(n)$ (bottom row) with radius $R$ (left), tidal deformability $\Lambda$ (middle), and Love number $k_2$ (right) for different NS masses as indicated obtained for a diverse set of nuclear models.}
\label{fig:models}
\end{figure}

It is interesting to compare our results, obtained from a Taylor
expansion parametrized around saturation density, with nuclear
models EoS. We use a dataset of unified EOS based on 24 Skyrme interactions and 26 relativistic mean-field nuclear parametrizations (some of them including a
transition to hyperonic matter at high density) \cite{Fortin18}. The
constraints $70<\Lambda_{1.4M_{\odot}}<580$ and
$M_{\text{max}}>1.97M_{\odot}$ are fulfilled by the following models:
BSk20, BSk21 \cite{Goriely2010}, BSk25, BSk26 \cite{Goriely2013}, SKa, SKb \cite{Kohler1976}, SkI4 \cite{Reinhard1995}, SkI6 \cite{Nazarewicz1996}, SkMP \cite{Bennour1989}, SKOp \cite{Reinhard1999}, SLy2, SLy9 \cite{Chabanat1995}, 
SLy230a \cite{Chabanat1997a}, SLy4 \cite{Chabanat1997b}.
In Fig.~\ref{fig:models} we show the variation with the density of the correlation coefficient between $P$, $E$, and $v_{s}$ and the NS properties $R$, $\Lambda$ and $k_2$ for different masses. When we compare the pressure
correlations in Fig.~\ref{fig:pressure} and in top panel of
Fig.~\ref{fig:models}, we notice that the main difference occurs at high
densities $n>0.6$~fm$^{-3}$, where the nuclear models show a stronger
correlation with all NS observables. 
We indicate in Table~\ref{tab:max_corr} the maximum
correlation obtained for $M=1.4M_{\odot}$ and the densities at which
they occur for the nuclear models. For reference, the correlation coefficient at the same densities obtained for our
EoSs from constrained meta-models are also given.
We see that the maximum correlations happen at nearly the same densities. The correlations of NS properties with the energy density
obtained for our set of EoSs from meta-models (Fig~\ref{fig:energy}) and those for
nuclear models (middle panel of Fig.~\ref{fig:models}) show overall similar trends. The fact that for our EoSs, the correlations are more suppressed at high 
densities may be due to the differences in the high density behavior.

\begin{table}[!htb]
\centering
\setlength{\tabcolsep}{10pt}
\renewcommand{\arraystretch}{1.1}
\begin{tabular}{ccccc}
  \hline \hline 
 Models&  &  & Nuclear & Meta\\      \hline
 &  & $n$ & corr & corr\\ 
  \hline
$P(n)$ & ${\Lambda_{1.4M_{\odot}}}$ & 0.320 & 0.992    & 0.891\\
$P(n)$ & ${R_{1.4M_{\odot}}}$       & 0.281 & 0.980    & 0.937\\
$e(n)$ & ${\Lambda_{1.4M_{\odot}}}$ & 0.544 & 0.978    & 0.914\\
$e(n)$ & ${R_{1.4M_{\odot}}}$       & 0.442 & 0.995    & 0.974\\
$v_s(n)$ & ${\Lambda_{1.4M_{\odot}}}$ & 0.219 & 0.987  & 0.870\\
$v_s(n)$ & ${R_{1.4M_{\odot}}}$     & 0.195 & 0.964    &0.844\\
   \hline \hline 
\end{tabular}
\caption{Densities of maximum correlations for the nuclear models and the correlation value at those densities for our set obtained from meta-models (last column). }
\label{tab:max_corr}
\end{table}

\subsection{Constraining the thermodynamical properties}

In the previous section, we saw that all thermodynamical quantities show
strong correlations with NS observables, specially the radius and the
tidal deformability, in some specific range of density. We now study
how to use these linear dependences to constrain the
thermodynamical quantities from future measurements of the radius and the
tidal deformability for a canonical NS, $M=1.4M_{\odot}$. \\

\begin{figure*}[!htb]
	\centering
	\includegraphics[width=0.46\columnwidth]{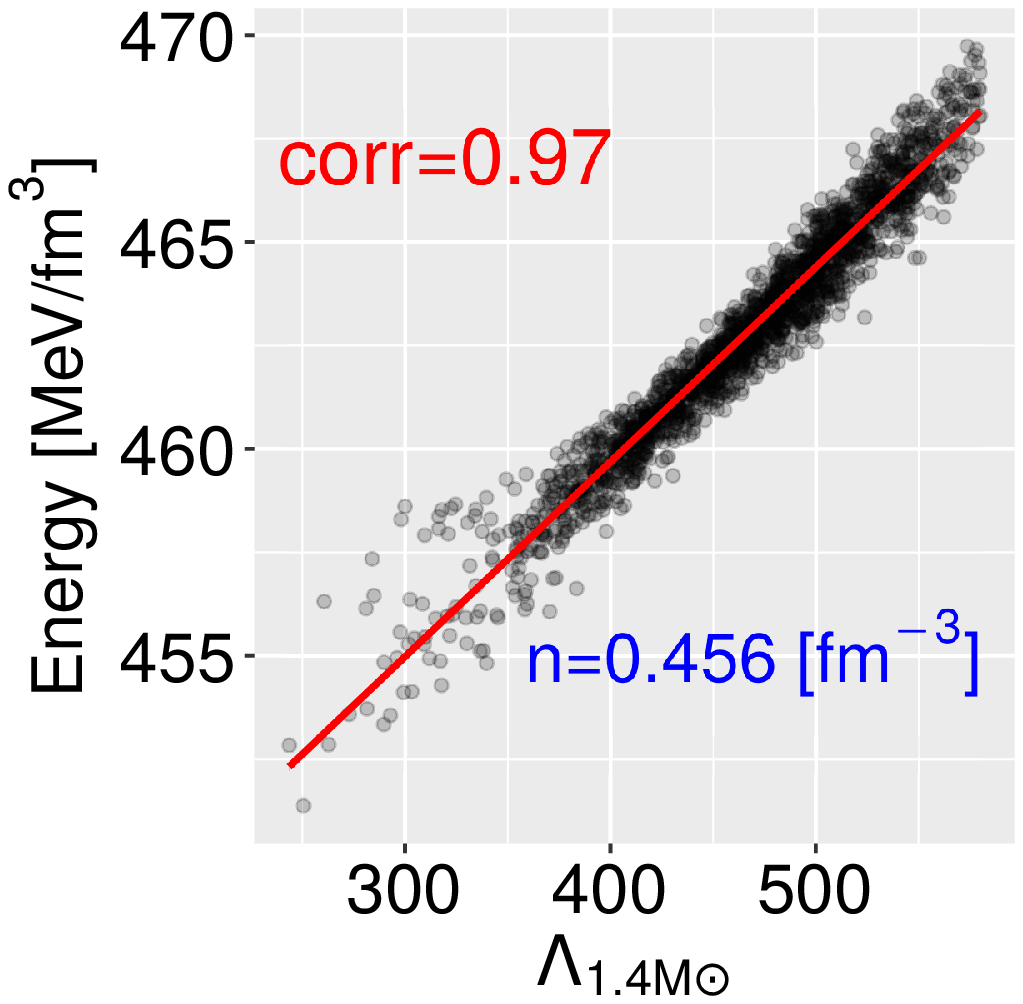}
	\includegraphics[width=0.46\columnwidth]{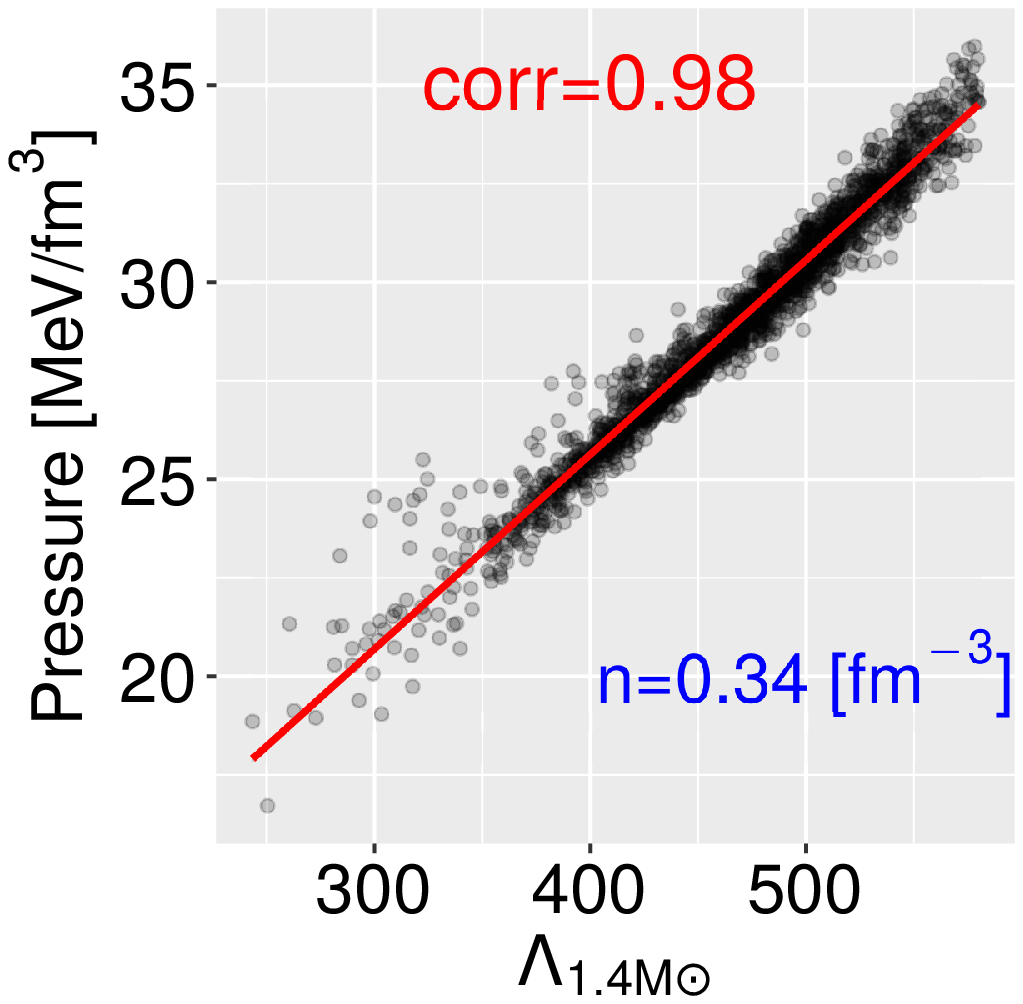}
	\includegraphics[width=0.46\columnwidth]{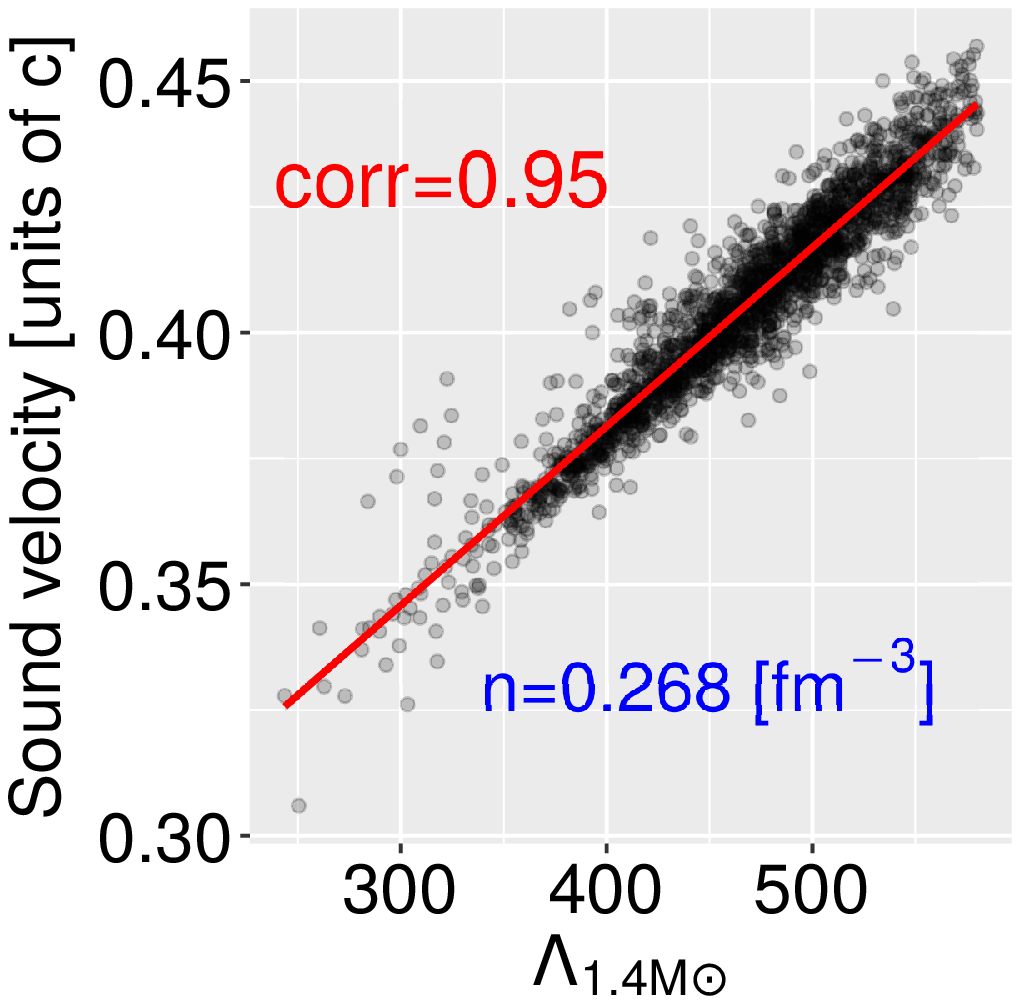}\\
	\includegraphics[width=0.46\columnwidth]{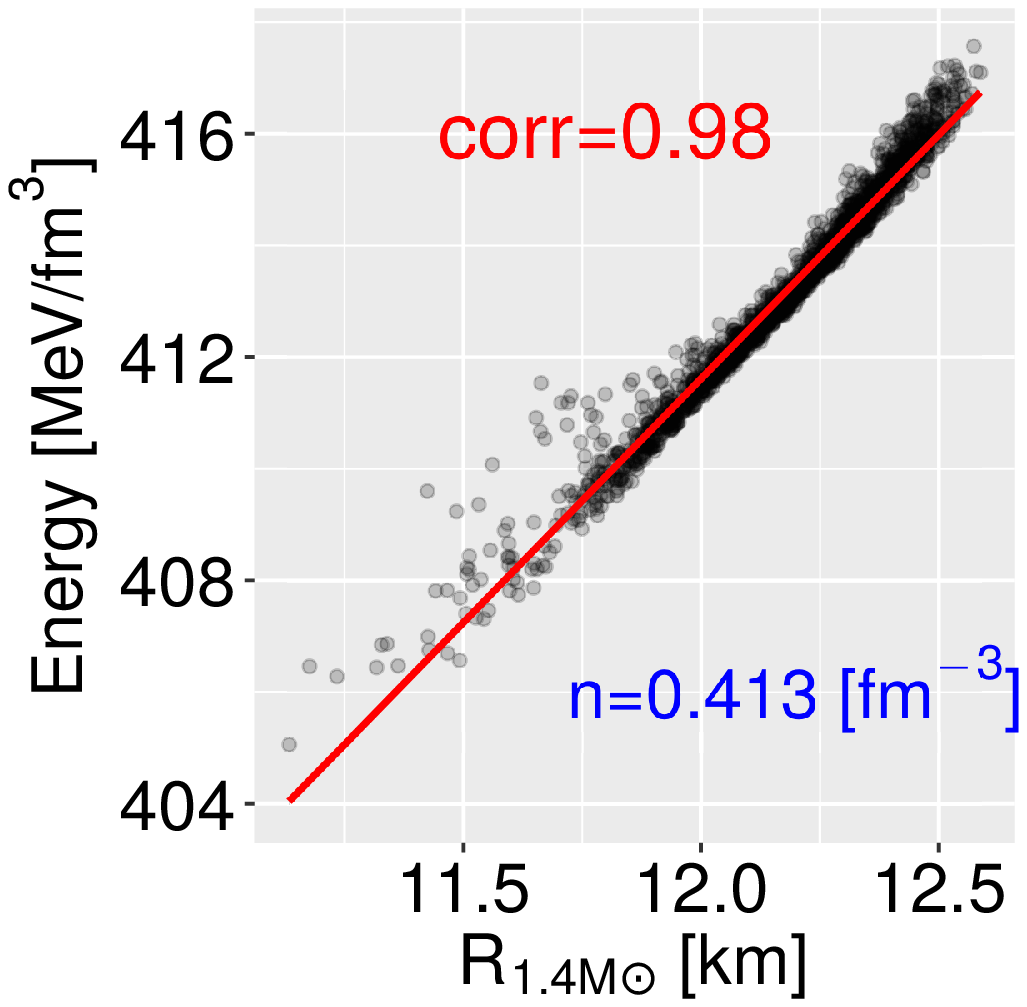}
	\includegraphics[width=0.46\columnwidth]{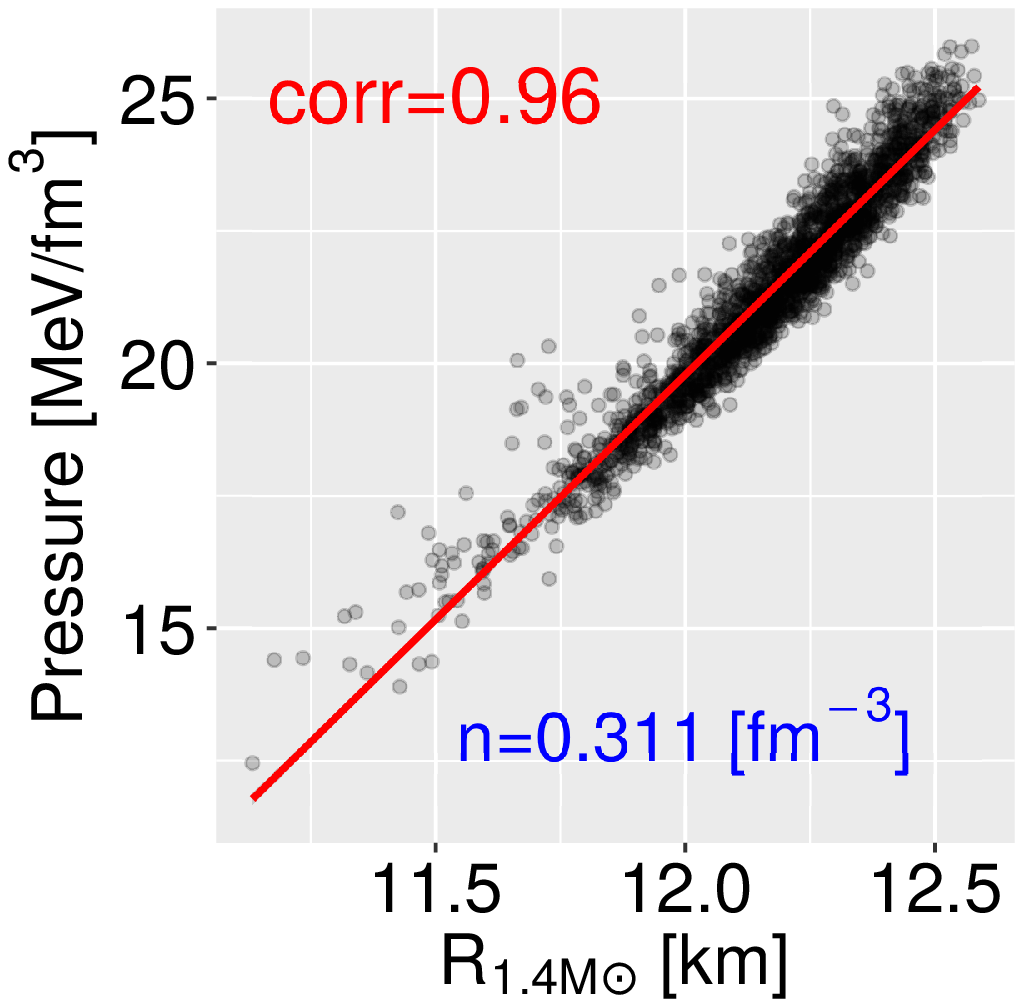}
	\includegraphics[width=0.46\columnwidth]{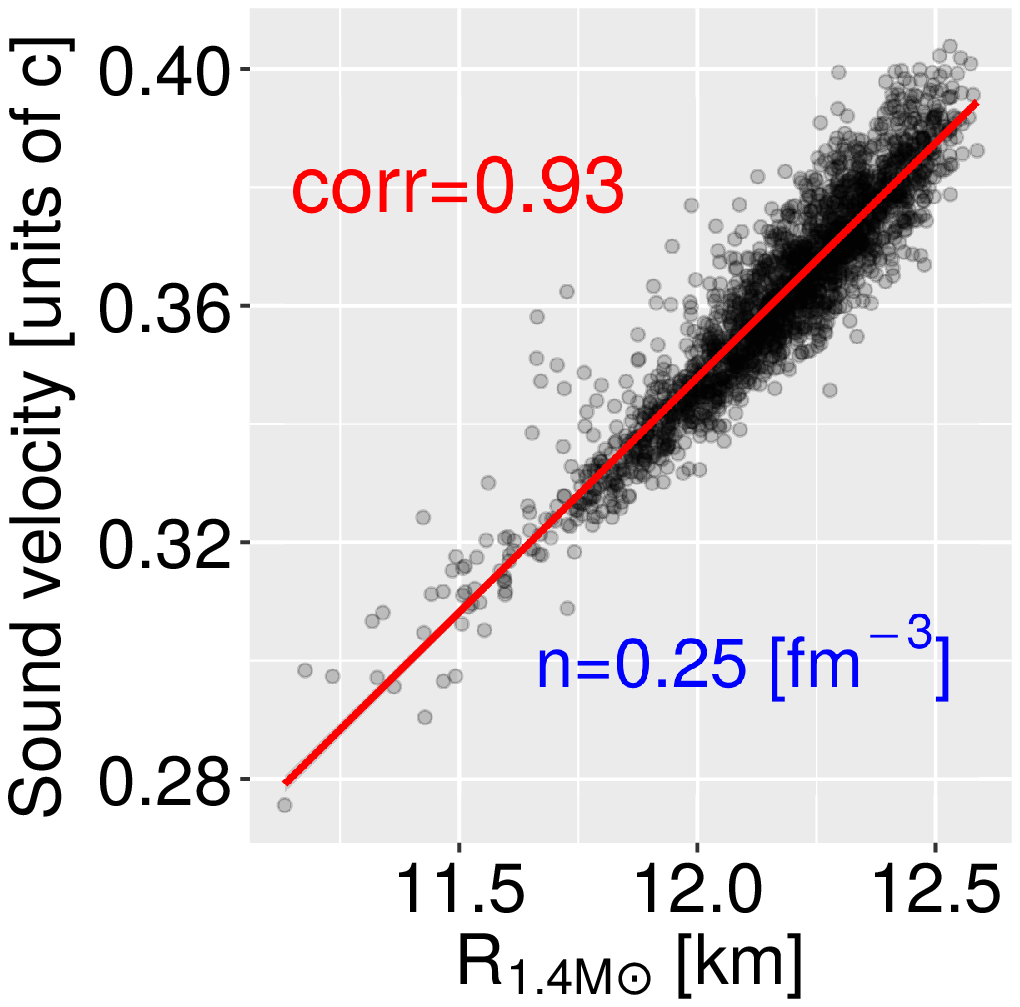}
	\caption{$\Lambda_{1.4 M_\odot}$ (top) and $R_{1.4 M_\odot}$ (bottom) as a 
 		  function of the energy density (left), the pressure (middle)
          and the speed of sound (right) for all meta models at the 
          densities corresponding to the maximum correlations. These 
          densities and correlation coefficient are indicated in each of the panels.}
	\label{corr}
\end{figure*}

Fig.~\ref{corr} shows the density at which the correlation is maximum
between both the tidal deformability (top panels) and radius (bottom panels) 
with the energy density (left), pressure (middle), and
sound velocity (right) of a 1.4$M_\odot$ star.  
The regression analysis is summarized in Table~\ref{tab:regressions} for three NS masses:  $1.0M_{\odot}$, $1.4M_{\odot}$ (shown in Fig.~\ref{corr}), and $1.8M_{\odot}$.
Recently, the first simultaneous determination of the radius and mass of a NS with the NICER mission 
	was obtained after the modeling the pulsating X-ray emission from the isolated millisecond pulsar PSR J0030+0451: $R\sim 11.5-14$ km for $M\sim1.4 M_\odot$ \cite{NICERa,NICERb,NICERc}.  If a more precise radius measurement becomes available, one would then be able to 
immediately get constraints on the EoS at three different densities: the
energy density at 0.413 fm$^{-3}$, the pressure at  0.311 fm$^{-3}$
and the sound velocity at  0.250 fm$^{-3}$. 
Similar constraints could also be derived from the observation of another NS, one of the primary targets PSR J0437$-$4715 with a mass 1.44 $\pm 0.07 M_\odot$
\cite{NICER,Watts16}. If besides also information
on the radius of the massive pulsar PSR J1614$-$2230 ($M=1.908\pm0.018M_\odot$)
 would be measured we would
be able to get further constraints on the EoS for  other different densities. 

From Table~\ref{tab:regressions}, using the maximum and the minimum $\Lambda_{1.4M_{\odot}}$ calculated from our EoS set (see Table~\ref{stat1}), we constrain the thermodynamical quantities at different densities 
\begin{eqnarray}
P(n&=&0.340\text{ fm}^{-3})=[17.82,34.56] \text{ MeV}/\text{fm}^3 \label{p}\\%
E(n&=&0.456\text{ fm}^{-3})=[452.22,468.22] \text{ MeV}/\text{fm}^3
  \label{e}\\
v_s(n&=&0.266\text{ fm}^{-3})=[0.325,0.446] \text{ units of c}. \label{vs}
\end{eqnarray}
Doing the same analysis for $R_{1.4M_{\odot}}$, using our sample max/min values, we get
\begin{eqnarray}
P(n&=&0.311\text{ fm}^{-3})=[11.66,25.27] \text{ MeV}/\text{fm}^3 \label{p}\\%
E(n&=&0.413\text{ fm}^{-3})=[403.98,416.78] \text{ MeV}/\text{fm}^3
  \label{e}\\
v_s(n&=&0.242\text{ fm}^{-3})=[0.278,0.395] \text{ units of c}. \label{vs}
\end{eqnarray}

These results are summarized in Table~\ref{tab:final_results}.
We also include, in the same Table, the constraints on $P$, $E$, and $v_s$ obtained from the LIGO/Virgo GW170817 analysis \cite{Abbott18}. 
Confidence intervals (on the marginalized posterior) for the pressure were determined in \cite{Abbott18}.
The 90\% credible level is
$P(n=0.311\text{ fm}^{-3})=[8.79,31.09] \text{ MeV}/\text{fm}^3$ and
$P(n=0.340\text{ fm}^{-3})=[12.01,41.39] \text{ MeV}/\text{fm}^3$, 
while the 50\% credible level is 
$P(n=0.311\text{ fm}^{-3})=[11.95,23.66] \text{ MeV}/\text{fm}^3$ and
$P(n=0.340\text{ fm}^{-3})=[16.37,32.00] \text{ MeV}/\text{fm}^3$.
Our results, Eqs. (8) and (11), are in good agreement even at the  50\% credible level.
We should note, however, that the EoSs used in \cite{Abbott18} have lower
$\Lambda_{1.4M_{\odot}}$ values, which are are not represented in our
dataset. Assuming that the correlations that we have obtained are model-independent, as the analysis of the nuclear models seems to suggest, 
one may use the regression analysis in Table~\ref{tab:regressions} to constrain the pressure at any given value of $\Lambda_{1.4M_{\odot}}$ (the same applies to $R_{1.4M_{\odot}}$). We get for $70< \Lambda_{1.4M_{\odot}}<580$, $P(0.340\text{ fm}^{-3})=[9.19,34.56] \text{ MeV}/\text{fm}^3$, which is close to the
interval determined in \cite{Abbott18}.
Furthermore, additional constraints on the energy density and sound velocity can be determined from $70<\Lambda_{1.4M_{\odot}}<580$.
	They are listed in Table~\ref{tab:final_results}.
 
\begin{table*}[!htb]
\centering
\setlength{\tabcolsep}{14pt}
\renewcommand{\arraystretch}{1.2}
\begin{tabular}{cccccc}
\hline \hline 
$n$     & $Q$                       & $Z$                      & Corr$[Q,Z]$ & $m$                               & $b$               \\ \hline
$0.275$ & \multirow{6}{*}{$P(n)$}   & $\Lambda_{1.0M_{\odot}}$ & $0.984$     & $(4.453\pm 0.017)\times 10^{-3}$  & $0.827 \pm 0.052$ \\
$0.340$ &                           & $\Lambda_{1.4M_{\odot}}$ & $0.982$     & $(49.389\pm 0.209)\times 10^{-3}$ & $5.898 \pm 0.098$ \\
$0.430$ &                           & $\Lambda_{1.8M_{\odot}}$ & $0.970$     & $(43.10\pm 0.23)\times 10^{-2}$   & $24.78 \pm 0.21$  \\ \\
$0.254$ &                           & $R_{1.0M_{\odot}}$       & $0.943$     & $(503.89\pm 3.86)\times 10^{-2}$  & $-49.04\pm 0.46$  \\
$0.311$ &                           & $R_{1.4M_{\odot}}$       & $0.962$     & $(942.68\pm 5.85)\times 10^{-2}$  & $-91.11\pm 0.70$  \\
$0.403$ &                           & $R_{1.8M_{\odot}}$       & $0.939$     & $(1885.9\pm 15.0)\times 10^{-2}$  & $-171.6\pm 1.8$   \\ \hline 
$0.377$ & \multirow{6}{*}{$E(n)$}   & $\Lambda_{1.0M_{\odot}}$ & $0.955$     & $(4.4\pm 0.0)\times 10^{-3}$     & $359.8 \pm 0.1$   \\
$0.456$ &                           & $\Lambda_{1.4M_{\odot}}$ & $0.973$     & $(47.4\pm 0.2)\times 10^{-3}$     & $440.9 \pm 0.1$   \\
$0.564$ &                           & $\Lambda_{1.8M_{\odot}}$ & $0.973$     & $(374.7\pm 1.9)\times 10^{-3}$     & $564.8 \pm 0.2$   \\ \\
$0.333$ &                           & $R_{1.0M_{\odot}}$       & $0.992$     & $(4583.9\pm 12.3)\times 10^{-3}$  & $271.3 \pm 0.1$   \\
$0.413$ &                           & $R_{1.4M_{\odot}}$       & $0.984$     & $(8823.4\pm 34.4)\times 10^{-3}$  & $306.8 \pm 0.4$   \\
$0.529$ &                           & $R_{1.8M_{\odot}}$       & $0.960$     & $(1628.0\pm 10.3)\times 10^{-2}$  & $354.5 \pm 1.3$   \\ \hline 
$0.209$ & \multirow{6}{*}{$v_s(n)$} & $\Lambda_{1.0M_{\odot}}$ & $0.960$     & $(4.908\pm 0.031)\times 10^{-5}$  & $0.160 \pm 0.001$ \\
$0.268$ &                           & $\Lambda_{1.4M_{\odot}}$ & $0.953$     & $(0.361\pm 0.002)\times 10^{-3}$  & $0.239 \pm 0.001$ \\
$0.346$ &                           & $\Lambda_{1.8M_{\odot}}$ & $0.934$     & $(1.918\pm 0.016)\times 10^{-3}$  & $0.360 \pm 0.001$ \\ \\
$0.192$ &                           & $R_{1.0M_{\odot}}$       & $0.921$     & $(66.371\pm 0.610)\times 10^{-3}$ & $-0.507\pm 0.007$ \\
$0.242$ &                           & $R_{1.4M_{\odot}}$       & $0.930$     & $(80.609\pm 0.694)\times 10^{-3}$ & $-0.604\pm 0.008$ \\
$0.323$ &                           & $R_{1.8M_{\odot}}$       & $0.898$     & $(93.086\pm 0.992)\times 10^{-3}$ & $-0.620\pm 0.012$ \\ \hline \hline 
\end{tabular}
\caption{Maximum correlations,  $\text{Corr}[Q,Z]$, between the
  thermodynamic properties, $Q=\{P(n), E(n), v_s(n)\}$, and the NS
  properties, $Z=\{\Lambda_{M_i},R_{M_i}\}$. We show the linear
  regression analysis for $Q=m\times Z+b$ at a fixed density
  $n$. 
    In the table $b$ has the units
    of $Q$ (MeV/fm$^3$ for $P$ and $E$ and $c$ for $v_s$) and $m$ has
    the units of $Z/Q$, i.e $Q^{-1}$ for $Z=\Lambda$ and  $km/Q^{-1}$ for $Z=R$.
    }
\label{tab:regressions}
\end{table*}

\begin{table*}
\centering
\setlength{\tabcolsep}{15pt}
\renewcommand{\arraystretch}{1.4}
\begin{tabular}{cccccccc}
\hline \hline
\multicolumn{2}{c}{\multirow{3}{*}{Bounds}}                & \multicolumn{2}{c}{\multirow{2}{*}{\begin{tabular}[c]{@{}c@{}}$P(0.340)$ \\ $[{\rm MeV~ fm}^{-3}]$\end{tabular}}} & \multicolumn{2}{c}{\multirow{2}{*}{\begin{tabular}[c]{@{}c@{}}$E(0.456)$ \\ $[{\rm MeV~ fm}^{-3}]$\end{tabular}}} & \multicolumn{2}{c}{\multirow{2}{*}{\begin{tabular}[c]{@{}c@{}}$v_s(0.268)$ \\ $[c]$\end{tabular}}} \\ 
\multicolumn{2}{c}{}                                       & \multicolumn{2}{c}{}                                                                                           & \multicolumn{2}{c}{}                                                                                           & \multicolumn{2}{c}{}                                                                             \\ \cline{3-8}
\multicolumn{2}{c}{}                                       & min                                                    & max                                                   & min                                                    & max                                                   & min                                             & max                                            \\ \hline

\multirow{2}{*}{$\Lambda_{1.4M\odot}$} 
& 243.53 – 579.93 & 17.82                                                  
& 34.56                                                 & 452.22                                                 & 468.22                                                & 0.325                                           & 0.446                                         
\\
&70 – 580 \cite{Abbott18}       
& 9.19                                                   & 34.56                                                 & 443.96                                                 & 468.23                                                & 0.262                                           & 0.446                                           \\
                                       \hline
                                       \multicolumn{2}{c}{\multirow{3}{*}{Bounds}}                & \multicolumn{2}{c}{\multirow{2}{*}{\begin{tabular}[c]{@{}c@{}}$P(0.311)$ \\ $[{\rm MeV~ fm}^{-3}]$\end{tabular}}} & \multicolumn{2}{c}{\multirow{2}{*}{\begin{tabular}[c]{@{}c@{}}$E(0.413)$ \\ $[{\rm MeV~ fm}^{-3}]$\end{tabular}}} & \multicolumn{2}{c}{\multirow{2}{*}{\begin{tabular}[c]{@{}c@{}}$v_s(0.242)$ \\ $[c]$\end{tabular}}} \\ 
\multicolumn{2}{c}{}                                       & \multicolumn{2}{c}{}                                                                                           & \multicolumn{2}{c}{}                                                                                           & \multicolumn{2}{c}{}                                                                             \\ \cline{3-8}
\multicolumn{2}{c}{}                                       & min                                                    & max                                                   & min                                                    & max                                                   & min                                             & max                                            \\ \hline
\multirow{2}{*}{$R_{1.4M\odot}$}       
                                       & 11.13 – 12.59   & 11.66                                                  & 25.27                                                 & 403.98                                                 & 416.78                                                & 0.278                                           & 0.395  \\
& 10.5 – 13.3  \cite{Abbott18}   & 5.74                                                   & 31.93                                                 & 398.40                                                 & 423.05                                                & 0.227                                           & 0.453                                         
\\ \hline \hline                                       
\end{tabular}
\caption{Constraints on pressure $P(n)$, energy density $E(n)$ and sound velocity $v_s(n)$
for $\beta$- equilibrium matter at density $n$(fm$^{-3}$) obtained from bounds on
tidal deformability and radius for neutron star with canonical mass $1.4M_\odot$ using
our EoSs. Constraints obtained from the LIGO/Virgo analysis are also shown (see text).}
\label{tab:final_results}
\end{table*}

\begin{figure}[!htb]
	\centering
	\includegraphics[width=1.0\columnwidth]{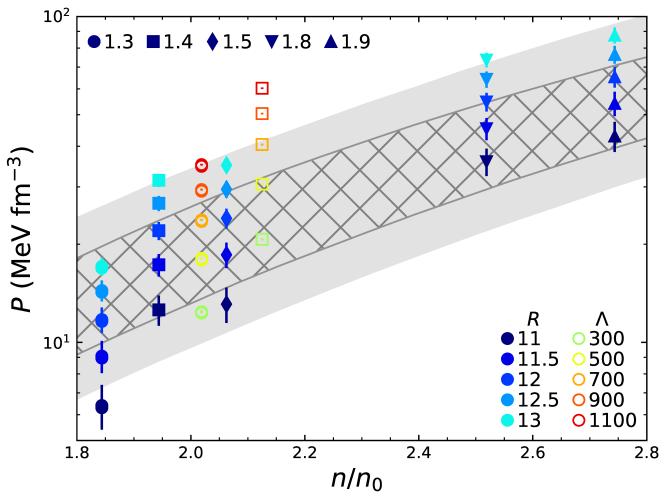}	
\caption{Constraints on the pressure obtained at different densities
(expressed in units of the nuclear saturation density) from measurements
of the radius (full symbols) and tidal deformabilities (empty symbols) of
NSs of different masses (shown by different types of symbols). The outer
(inner) gray region is the 90\% credible level (50\% credible level) from
\cite{Abbott18}. See text for details.} \label{fig:Reconst} \end{figure}

In Fig.~\ref{fig:Reconst} we show the constraints on the pressure at
different densities which  one can obtain from measurements of the radius and
tidal deformability for  NSs of various masses. For the radius, we adopt
the  masses of $M=1.3, 1.4$ and $1.5M_\odot$, which covers the range of
values for most observed NSs in particular the NICER targets PSR J0030+0451 and  PSR
J0437$-$4715 and $M=1.8$ and $1.9M_\odot$ to explore the consequence of
the radius determination of a massive NS such as PSR J$1614-2230$. For
the tidal deformability we restrict ourselves to $M=1.3$ and $1.4M_\odot$
as this corresponds to the mass range of NSs  currently observed in
a binary with another NS \cite{Alsing18}. The vertical error bars
(hardly visible)  for constraints from the tidal deformabilties
account for the errors in the determination of the slope and the
interception in the linear regression with the pressure (parameters
$m$ and $b$ in Table~\ref{tab:regressions}). For medium-range masses
$M\sim1.4M_\odot$ the future determination of the tidal deformability will
allow to constrain the pressure at densities $\sim10\%$ larger than that  for  the
radius. In addition, measuring the radius of a massive NS would allow us
to put limits at the pressure for larger densities. Thus observational
constraints on NS properties from multi-messenger astrophysics will
enable us to infer the properties of NS matter in the range $1-3n_0$
which we can not probe in terrestrial laboratories.

\section{Conclusions}

We found strong correlations between the thermodynamical properties,
i.e., the pressure, energy density and sound velocity with the radius
and the tidal deformability of NS over a wide range of masses.  These
correlations were obtained from a set of EoS parametrized by a Taylor
expansion around the saturation density. Similar correlations are confirmed 
using EoS obtained from nuclear models, indicating that they are model-independent.
It is shown that for a given NS mass, there is always a
density where the tidal deformability and the radius are highly correlated
with the pressure, energy density, and speed of sound. A single
determination of the tidal deformability $(\Lambda_{M_i})$ or radius
$(R_{M_i})$ of a NS of mass $M_i$ allows to constrain the thermodynamic properties at three distinct but close densities. For a $1.4M_{\odot}$
NS, the pressure could be constrained at $n\approx 2n_0$, the energy
density at $n\approx 2.7n_0$, and the sound velocity at $n\approx 1.7n_0$.

We show that the radius and  tidal deformability of NSs for different
masses are  strongly  correlated with thermodynamic variables of EoS at
supra-saturation densities in the range of   $n\approx 1-3n_0$.
The precise  measurement of the radius of the pulsars PSR J0030+0451 and  PSR J0437$-$4715 with a
mass $\sim1.4 M_\odot$ by the NICER mission would allow us to get
immediately information on the EoS at three different densities, the
pressure at  $\sim 2n_0$, the energy density at $\sim2.5n_0$, and
the sound velocity at  $\sim 1.5n_0$.  Complementing this information
with the further radius measurement, for example, of  PSR J1614-2230 with
$M\simeq1.9M_\odot$ and constraints on the tidal deformability of various
NSs from the observations of GW from merging binary NS systems would set
very stringent constraints on the EoS at supra-saturation densities.  
It may be pointed out that if a phase transition to a quark phase takes place, this will occur at densities close to or beyond $3n_0$, therefore it will may not significantly affect the correlations obtained in the present work, as they are connected to the hadronic phase.

\section*{ACKNOWLEDGMENTS}
This article is based upon work partially funded by  COST Action CA16214   PHAROS  supported by COST (European Cooperation in Science and Technology),
and by the FCT (Portugal) Projects No.UID/FIS/04564/2019 and POCI-01-0145-FEDER-029912 and the Polish National Science
Centre (NCN) under the grant No. 2017/26/D/ST9/00591.
T.M acknowledge the Council of Scientific and Industrial Research (CSIR), Human Resource Development, Govt. of India for 
the support of International Travel Grant TG/10542/19-HRD.

\end{document}